\documentclass[showpacs,aps,pra,amsmath,amssymb,twocolumn]{revtex4}
\usepackage{graphicx,graphics}
\usepackage{dcolumn}
\usepackage{bm,color}
\usepackage{txfonts,hyperref}
\usepackage{amsmath}
\usepackage{epstopdf}
\newcommand{\ket}[1]{\left\vert#1\right\rangle}
\newcommand{\bra}[1]{\left\langle#1\right\vert}

\newcommand{\eq}{Eq.~}
\newcommand{\eqs}{Eqs.~}
\newcommand{\fig}{Fig.~}

\newcommand{\cf} {cf.~}
\newcommand{\ug} {\!=\!}

\newcommand{\piu} {\!+\!}
\newcommand{\meno} {\!-\!}

\newcommand{\eg} {e.g.~}
\newcommand{\etal} {{\it et al.}}

\newcommand{\rref} {Ref.~}

\newcommand{\vs} {vs.~}

\newcommand{\ibid}{{\it ibid.~}}  
\providecommand{\av}[1]{\left\langle #1\right\rangle}

\providecommand{\ket}[1]{|#1\rangle}
\providecommand{\bra}[1]{\langle#1|}
\providecommand{\kebra}[2]{\ket{#1}\!\bra{#2}}
\providecommand{\sprod}[2]{\langle#1|#2\rangle}

\begin{document}
 
 
 
\title{Dynamics of spontaneous emission in a single-end photonic waveguide}

\author{Tommaso Tufarelli \mbox{$^{1}$}}
\author{Francesco Ciccarello\mbox{$^{2,3}$}}
\author{M. S. Kim\mbox{$^{1}$}}
 
\affiliation{\\ \mbox{$^{1}$QOLS, Blackett Laboratory, Imperial College London, SW7 2BW, UK}\\
\mbox{$^{2}$}Scuola Normale Superiore, Piazza dei Cavalieri, 7, I-56126 Pisa, Italy\\
\mbox{$^{3}$}{NEST, Istituto Nanoscienze-CNR and Dipartimento di Fisica, Universit$\grave{a}$  degli Studi di Palermo, via Archirafi 36, I-90123 Palermo, Italy}}
\begin{abstract}
We investigate the spontaneous emission of a two-level system, \eg an atom or atomlike object, coupled to a  single-end, {\it i.e.}, semi-infinite, one-dimensional photonic waveguide such that one end behaves as a perfect mirror while light can pass through the opposite end with no back-reflection. Through a quantum microscopic model we show that such geometry can cause non-exponential and long-lived atomic decay.  Under suitable conditions, a bound atom-photon stationary state appears in the atom-mirror interspace so as to trap a considerable amount of initial atomic excitation. Yet, this can be released by applying an atomic frequency shift causing a revival of photon emission. The resilience of such effects to typical detrimental factors is analyzed.
\end{abstract}

\maketitle
\noindent
 
\section{Introduction}  A major, if not distinctive, line in quantum electrodynamics (QED) is to study how geometric constraints affect the interaction between atomic systems and the electromagnetic (EM) field. On the one hand, this can bring a deeper insight into the related physics. On the other hand, phenomena that spontaneously do not occur in Nature can become observable this way.
Spontaneous emission (SE) is an elementary process in QED. One normally associates this with an exponential decay of a quantum emitter (QE) to its ground state accompanied by an irreversible release of energy to the EM vacuum (we will often use the term ``atom" to refer to the QE even though this needs not be necessarily an actual atom). 
However, free-space SE can be significantly affected -- even in its qualitative features -- by introducing geometric constraints forcing the EM field within a certain region of space \cite{purcell, meschede,cqed} or a lattice structure (see \eg \cite{bandgap}).
Cavity-QED has embodied for a long time the traditional test-bed for investigating such effects. Nowadays, growing technologic capabilities to effectively confine the EM field within less-than-three dimensions and make it interact with a small number of atoms are opening the door to yet unexplored areas of QED. In particular, a variety of experimental implementations of one-dimensional (1D) photonic waveguides coupled to few-level systems have been developed. 
These include photonic-crystal waveguides with defect cavities \cite{pc}, optical or hollow-core fibers interacting with atoms \cite{fibers}, microwave transmission lines coupled to superconducting qubits \cite{wallraf}, semiconducting (diamond) nanowires with embedded QDs (nitrogen vacancies) \cite{nws,gerard,delft} or plasmonic waveguides coupled to QDs or nitrogen vacancies \cite{plasmons} (see \rref \cite{sorensen} for a more comprehensive list). Interestingly, even free-space setups employing tightly focused photons have the potential to embody effective 1D waveguides \cite{freespace}.
 
In most cases, the number (even at the level of a single unity) and positioning of such atomlike objects can be accurately controlled. Besides major applicative concerns to study such 1D systems (\eg some of them can work as highly-efficient single-photon sources) these developments are fostering a renewed interest in their fundamental quantum optical properties. {Peculiar effects can take place, such as giant Lamb shifts \cite{agnellone} or the ability of an atom to perfectly reflect back an impinging resonant photon due to the destructive interference between spontaneous and stimulated emission \cite{shen-fan}. The latter effect is at the heart of attractive applications such as single-photon transistors \cite{lukin} and atomic light switches \cite{switch}.}
 
To capture certain features of their physical behavior, it often suffices to model 1D photonic waveguides as {\it endless}. In reality, of course, one such structure is terminated on both sides, each end typically lying at the junction between the waveguide itself and a solid-state or air medium. Thereby, light impinging on either end always undergoes some partial back-reflection owing to refractive-index mismatch. Yet, mostly prompted by the wish to realize efficient single-photon sources, 
latest technology is now attaining the fabrication of {\it single-end}, quasi-1D structures. For instance, this can be achieved by tapering the waveguide towards one end so as to make this almost transparent, while the opposite end is joined to an opaque medium \cite{gerard,delft}.
The system thus behaves as being semi-infinite. Equivalently, it can be regarded as an infinite waveguide with a perfect mirror (embodied by the opaque end). Given this state of the art, a thorough knowledge of the emission process of an atom in such a configuration is topical. While the analogous problem in 3D space has been studied extensively \cite{meschede}, first insight into the SE of a QE in a semi-infinite 1D waveguide has been acquired only recently through semiclassical \cite{jap,friedler} and quantum models \cite{sung}.  
Unlike the 2D or 3D cases, the peculiarity of this 1D setup is that the entire amount of radiation emitted by the atom and back-reflected by the mirror is constrained to return to the emitter, and hence has a significant chance to re-interact with it. As is typical in such circumstances, due to multiple reflections, the atom-mirror optical path length becomes crucial and resonances are introduced in the system. 
This is witnessed by very recent studies (although not focusing on SE), where the waveguide termination was shown to drastically benefit microwave-single-photon detection \cite{solano}, atomic inversion schemes \cite{chen} and processing of quantum information encoded in QEs \cite{gate} and photons \cite{valente}. 
 
In \rref\cite{sung}, through a stationary approach suited to atom-photon 1D scattering \cite{shen-fan} it was shown that quasibound states can emerge in a single-end waveguide coupled to an atom. Indeed, it is known that an atom can behave as a perfect mirror itself \cite{shen-fan,freespace,switch}, hence effective cavities with atomic mirrors can be formed \cite{supercavity}.
 
Here, we study the SE of a two-level system coupled to a semi-infinite waveguide through the analysis of a fully quantum model and a purely dynamical approach. We find that non-exponential emission with long time tails occurs in general. This can feature photon-reabsorption signatures and even excitation trapping, where the latter means that full atomic decay to the ground state is inhibited due to the emergence of an atom-photon bound state. We explain such effects in detail and illustrate the corresponding output light dynamics, thanks to a non-perturbative analysis of the system's time evolution, resulting in a closed delay differential equation governing the entire SE process.
 
The present paper is structured as follows.  In Section \ref{model}, we introduce our model and explain the method we used to tackle the SE dynamics. Some assumptions that we make are justified. In Section \ref{SED}, after working out the delay differential equation governing the atomic excitation time evolution, some typical examples of the entailed dynamics are shown. In particular, we illustrate the inhibition of a full atomic decay to the ground state. In Section \ref{bound}, this peculiar effect is demonstrated by working out analytically the excitation amplitude at large times. We also provide a physical explanation by showing that, correspondingly, within the mirror-atom interspace a bound state is formed, whose overlap with the initial state matches the asymptotic excitation amplitude. In Section \ref{out-field-sec}, we illustrate the dynamics of the output light exiting the waveguide and show an interesting method to induce an emission revival corresponding to a full release of the trapped excitation. In Section \ref{resilience}, we analyze how resilient are these phenomena to typical detrimental effects occurring in this type of setups. In Section \ref{conc}, we finally draw our conclusions. The work ends with three Appendixes, where some technical details are supplied.
 
\begin{figure}
\begin{center}
\includegraphics[width=0.65\linewidth]{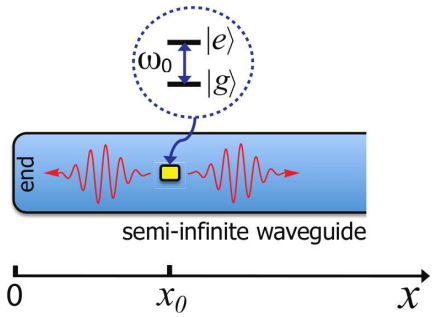}\end{center}\vspace{-.3cm}
\caption{(Color online) Setup. A semi-infinite waveguide, whose end lies at $x\ug0$, coupled to a two-level system at $x\ug x_0$. \label{Fig1}}
\end{figure} 
 
\section{Model and approach} \label{model} We consider a 1D semi-infinite waveguide along the $x$-axis, whose only termination lies at $x\ug0$. The waveguide is coupled at $x\ug x_0$ to a two-level atom, whose ground and excited states $\ket g$ and $\ket e$ have a frequency splitting $\omega_0$.
Thus $x_0$ is the distance between the atom and the waveguide end, the latter behaving as a perfect mirror. We sketch the entire setup in \fig1. The waveguide supports a continuum of electromagnetic modes, each with associated wave vector $k$,  frequency $\omega_k$ and annihilation (creation) operator  $\hat{a}_k$ ($\hat{a}^\dagger_k$), obeying the bosonic commutation rule $[\hat{a}_k,\hat{a}_{k'}^\dagger]\ug \delta(k\meno k')$. In the case of an infinite waveguide, for each $k\!>\!0$ two orthogonal standing modes are possible with spatial profiles $\propto\! \cos(kx)$ and $\propto\!\sin(kx)$, respectively. In our case, given that the waveguide terminates at $x\ug0$ only the sine-like modes are to be accounted for. Thereby, the atom is dipole-coupled to mode $k$ with strength $g_k\!\propto\!\sin(kx_0)$. 
By neglecting counter-rotating terms, the Hamiltonian reads
\begin{equation}
\hat{H}\ug\omega_0\kebra{e}{e}\piu\!\int_0^{k_c}\!\!{\rm d}k\;\omega_k\hat a^\dagger_k\hat a_k\piu\int_0^{k_c}\!\!{\rm d}k\left(g_k\, \hat\sigma_+\hat a_k\piu\textrm{H.c.}\right),\label{H}
\end{equation}
where $\hat\sigma_+\ug\hat\sigma_-^\dag\ug\kebra{e}{g}$ and $k_c$ stands for a cut-off wave vector depending on the specific waveguide. 
The total number of excitations is conserved since $[\hat{H},\kebra{e}{e}\piu\int\!{\rm d}k\;\hat a_k^\dagger \hat a_k]\ug0$. As we will focus on the atomic SE, the dynamics occurs entirely within the one-excitation sector of the Hilbert space. Thus, at time $t$ the wave function is of the form
\begin{equation}
\ket{\Psi(t)}\ug\varepsilon(t)\ket e\!\ket 0\piu\ket{g}\!\int\!{\rm d}k\; \varphi(k,t) \,a^\dagger_k\!\ket{0},\label{generic-state}
\end{equation}
where $\ket 0$ is the field vacuum state, $\varepsilon(t)$ is the atomic excitation probability amplitude and $\varphi(k,t)$ is the field amplitude in the $k$-space (the normalization condition $|\varepsilon(t)|^2\piu\int{\rm d}k\;|\varphi(k,t)|^2\ug1$ holds). Using \eqs(\ref{H}) and (\ref{generic-state}) and the bosonic commutation rules, the time-dependent Schr\"{o}dinger equation $\partial_t{\ket\Psi}\ug-i\hat{H}\ket\Psi$ yields the coupled differential equations:
\begin{align}
\dot \varepsilon(t)&\ug-i\omega_0\varepsilon(t)-i\!\int_0^{k_c} \!\!{\rm d}k\; g_k\,\varphi(k,t),\label{epsilon}\\
\partial_t \varphi(k,t)&\ug-i\omega_k\varphi(k,t)-ig_k^*\varepsilon(t)\label{phik}.
\end{align}
 
In line with standard approaches for tackling similar systems \cite{shen-fan, sorensen}, we shall make two main assumptions. First, the photon dispersion relation can be linearized around the atomic frequency as $\omega_k\!\simeq\! \omega_0\piu \upsilon (k\meno k_0)$, where $\upsilon\ug\left.{{\rm d}\omega}/{{\rm d}k}\right|_{k\ug k_0}$ is the photon group velocity and $k_0$ is such that $\omega_{\kappa_0}\ug \omega_0$.  
Moreover,  to simplify our calculations we approximate the integral bounds as $\int_0^{k_c} {\rm d}k\!\rightarrow\!\int_{-\infty}^\infty {\rm d}k$. These approximations, including the exclusion of the counter-rotating terms mentioned earlier, are valid because we will focus on processes where only a narrow range of wave vectors around $k\!\ug \! k_0$ is involved. Hence, wave vectors which are far from $k_0$ (including $k\!<\!0$ and $k\!>\!k_c$ that are unphysical) have negligible effect. In the following, we will set $g_k\ug \sqrt{\Gamma \upsilon/\pi}\sin{kx_0}$, where $\Gamma$ is the atomic SE rate if the waveguide were infinite (no mirror). This assumption will be justified {\it a posteriori} shortly. 

\section{Spontaneous emission dynamics}  \label{SED}Next, we study the system's dynamics when the atom and field are initially in $\ket e$ and $\ket 0$, respectively. The initial conditions thus read $\varepsilon(0)\ug 1$ and $\varphi(k,0)\ug 0$ for any $k$. We start by removing the central frequency $\omega_0$ from \eqs(\ref{epsilon}) and \eqref{phik}, via the transformation $\varepsilon(t)\!\to\!\varepsilon(t) {\rm e}^{-i\omega_0t},\varphi(k,t)\!\to\!\varphi(k,t){\rm e}^{-i\omega_0t}$. \eq\eqref{phik} is thus integrated in terms of the function $\varepsilon(t)$ as $\varphi(k,t)\ug-i\sqrt{\Gamma\upsilon/\pi}\sin{kx_0}\int_0^t\!{\rm d}s \,e^{i\upsilon(k-k_0)(s-t)}\varepsilon(s)$. Replacing this into \eq(\ref{epsilon}) gives
\begin{equation}\label{DDE0}
\dot\varepsilon(t)\ug \meno\tfrac{\Gamma\upsilon}{\pi}\!\!\int_0^t\!\!{\rm d}s\;\varepsilon(s)e^{\!-\!i \upsilon k_0 (s-t)}\!\!\int\!\!{\rm d}k\sin^2(k x_0)e^{i \upsilon k (s-t)}.
\end{equation}
The integral over $k$ is easily calculated as a linear combination of $\delta(s-t\pm t_d)$ and $\delta(s-t)$, where $t_d\ug 2x_0/\upsilon$ is the time taken by a photon to travel from the atom to the waveguide end and back [see \fig1]. 
Once this is used to carry out the integration over $s$, we end up with a \textit{delay differential equation} (DDE) for $\varepsilon(t)$ with associated {\it time delay} $t_d$:
\begin{align}
\dot\varepsilon(t)\ug -\tfrac{\Gamma}{2}\,\varepsilon(t)+\tfrac{\Gamma}{2}\, {\rm e}^{i\phi}\varepsilon(t-t_d)\theta(t-t_d)\,\,,\label{DDE}
\end{align}
where $\theta(t)$ is the Heaviside step function while the {\it phase} $\phi\ug 2k_0x_0$ is the optical length of twice the atom-mirror path [see \fig1]. DDEs typically occur in problems where retardation effects are relevant such as in the case of two distant QEs in free space \cite{dde} and a single emitter embedded in a dielectric nanosphere \cite{dde2}. A similar equation was recently obtained in \rref\cite{solano} by working in real space.
The first term on the right-hand side of Eq.~\eqref{DDE} describes a standard damping at a rate $\Gamma$. The second term, instead, indicates that atomic re-absorption of the emitted photon can occur at times $t\!\ge\! t_d$. { At earlier times, such re-absorption term is null since the photon has not yet performed a round trip between the atom and the mirror}. Also, it vanishes in the limit $t_d\!\to\!\infty$ since the waveguide then effectively becomes infinite and, as expected, the atom undergoes standard, namely fully irreversible, SE at a rate $\Gamma$ according to $|\varepsilon(t)|^2\ug e^{-\Gamma t}$. This justifies our parametrization of the coefficients $g_k$, introduced at the end of the previous section.
 
Eq.~\eqref{DDE} can be solved iteratively by partitioning the time axis into intervals of length $t_d$. By proceeding similarly to \rref \cite{dde} we obtain its explicit solution as
\begin{equation}
\varepsilon(t)={\rm e}^{\meno\frac{\Gamma}{2}t}\sum_{n}\frac{1}{n!}\left(\tfrac{\Gamma}{2}{\rm e}^{i\phi+\frac{\Gamma}{2}t_d}\right)^n(t-nt_d)^n\theta(t-nt_d),
\end{equation}
where the effect of multiple reflection and re-absorption events is witnessed by the presence of the Heaviside step functions (they add a new contribution to the sum at the end of each photon round trip).
\begin{figure}
\begin{center}
\includegraphics[width=1\linewidth]{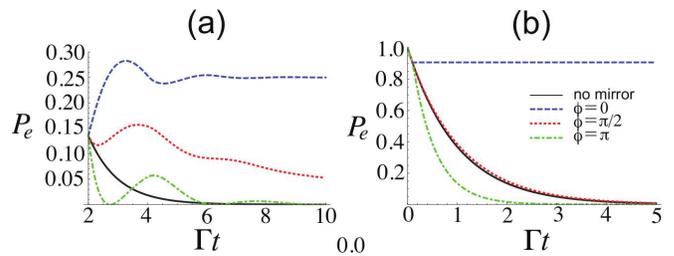}\end{center}\vspace{-.3cm}
\caption{(Color online) Atomic excitation probability $P_e(t)\ug|\varepsilon(t)|^2$ \vs time (in units of $1/\Gamma$) in the cases $\Gamma t_d\ug 2$ (a) and $\Gamma t_d\ug 0.1$ (b) and for $t_d\ug \infty$ (i.e., no mirror; solid black line), $\phi\ug 2n\pi$ (blue dashed line), $\phi\ug \pi/2\piu 2n\pi$ (red dotted) and $\phi\ug (2n+1)\pi$ (green dash-dotted). In (a), only the range $t\geq t_d$ is shown: at earlier times the behavior does not depend on $\phi$ and is the continuation of the black solid line.  \label{Fig2}}
\end{figure} 
In Fig.~2, we plot the time evolution of the atom excitation probability $P_e(t)\ug |\varepsilon(t)|^2$ for different values of $\phi$ for $\Gamma t_d\ug 2$ (a) and $\Gamma t_d\ug0.1$ (b). In either case, the expected purely exponential decay occurring with an infinite waveguide (i.e., the no-mirror case) is displayed for comparison. Such behavior clearly takes place even in the present setup as long as $t\!<\! t_d$ (independently of $\phi$ and $t_d$). As soon as $t\geq  t_d$, however, the presence of the mirror starts  affecting the atom in a way that the dynamics is now strongly dependent on $\phi$ and $ t_d$. For $\Gamma  t_d$ of the order of one, such as in \fig2(a), the behavior of the atomic population can deviate sensibly from an exponential decay: it exhibits one or more peaks of partial atomic re-excitation and, eventually, a monotonic decay. The phase $\phi$ affects both the positions of such re-excitation peaks and the long-time behavior of $P_e(t)$  [see Fig.~2(a)]. When instead $\Gamma t_d\ll1$, such as in \fig2(b), $P_e(t\!\ge\!t_d)$ drops monotonically with the phase $\phi$ simply affecting the atom's average lifetime. Indeed, in such regime the solution to Eq.~\eqref{DDE} can be approximated as (see Appendix \ref{derivation-eps-app})
\begin{align}\label{eps-app}
\varepsilon(t)\simeq{\rm e}^{-\tfrac{\Gamma}{2}t}\theta(t_d-t)+{\rm e}^{-\tfrac{\Gamma t_d}{2}}\left(\tfrac{1+{\rm e}^{i\phi}\tfrac{\Gamma t_d}{2}}{1+\tfrac{\Gamma t_d}{2}}\right)^{\tfrac{t-t_d}{t_d}}\!\!\!\theta(t-t_d)\,\,,
\end{align}
up to an irrelevant phase factor. The corresponding $|\varepsilon(t)|^2$ decays monotonically since $|{1+{\rm e}^{i\phi}\tfrac{\Gamma t_d}{2}}|\!\le\!|{1\piu\tfrac{\Gamma t_d}{2}}|$.
 
\section{Atom-photon bound state}  \label{bound}An interesting feature emerges for $\phi\ug 0$. Fig.~2 indeed shows that, regardless of $\Gamma t_d$, such optical path length inhibits a full excitation decay of the atom on the considered timescales. Indeed, it can be shown that the atom holds a significant amount of excitation even in the limit $t\!\to\!\infty$. To show this, we take the Laplace transform (LT) of Eq.~\eqref{DDE} and solve the resulting algebraic equation. This yields\vspace{-.2cm}
\begin{align}
\tilde\varepsilon(s)\ug \frac{1}{s+\tfrac{\Gamma}{2}(1-{\rm e}^{i\phi-s t_d})},
\end{align}
where $\tilde\varepsilon(s)$ is the LT of $\varepsilon(t)$. Using the final value theorem, we find the long-time limit \cite{nota-1order}
\begin{eqnarray}
\varepsilon(t\!\to\!\infty)\ug \lim_{s\rightarrow 0} [s\tilde{\varepsilon}(s)] \ug\left\{\begin{array}{ll}
\left(1+\tfrac{\Gamma t_d}{2}\right)^{-1}&\, \textrm{for}\;\;\phi\ug 2n\pi\vspace{.1cm}\\
\,\,\,\,\,\,\,0&\,\textrm{for}\;\;\phi\!\neq\! 2n\pi.
\end{array}\right.\label{FVT}
\end{eqnarray} 
In the former case note that, in particular, the asymptotic excitation increases when $t_d$ is reduced witnessing the crucial presence of the mirror. The lower $\Gamma$ the more significant is the increase, i.e., the less uncertain is the atomic emitted light wavelength the more pronounced is the effect suggesting an interference-like mechanism behind the phenomenon. Such inhibition of spontaneous emission can be interpreted as due to a destructive interference between the different paths that the emitted photon can take to exit the waveguide, or equivalently, between the probability amplitudes of emitting the photon at two different times. It is indeed the signature of a metastable {\it bound state} established between the atom and the photonic environment. The emergence of atom-photon bound states has been demonstrated in other different scenarios such as gapped photonic crystals \cite{bandgap} and super-Ohmic baths \cite{nonmarkovianeffect}.
 
If existent, an atom-photon bound state $\ket{\Psi_b}\ug \varepsilon_b \ket{e\,0}\piu \int\! {\rm d}k\,\varphi_b(k) \ket{g}\hat{a}_k^\dag\ket{0}$ must fulfill the normalization condition $\bra{\Psi_b}\Psi_b\rangle\ug1$ and the time-independent Schr\"{o}dinger equation $\hat{H}\ket{\Psi_b}\ug E\ket{\Psi_b}$. With the replacement $\partial_t\!\to\!-iE$ in Eqs.~\eqref{epsilon} and \eqref{phik}, once the rescaled energy parameter $q\ug(E-\omega_0)/\upsilon$ is introduced, we end up with the following equations for the atomic and field amplitudes
\begin{align}
q\,\varepsilon_b\ug \sqrt{\Gamma /(\upsilon\pi)}\!\int\!\!{\rm d}k\sin(k x_0)\varphi_b(k),\label{epsilon2}\\
[q\meno(k\meno k_0)]\varphi_b(k)\ug \sqrt{\Gamma/( \upsilon\pi)}\sin(k x_0)\varepsilon_b\,.\label{phik2}
\end{align}
Solving Eq.~\eqref{phik2} for $\varphi_b(k)$, we obtain that the squared norm of the bound state is given by $\bra{\Psi_b}\Psi_b\rangle\ug[1\piu \Gamma /(\pi\upsilon) \int\!{\rm d}k\sin^2(kx_0)/(k\meno k_0\meno q)^2]|\varepsilon_b|^2$. Calculating the integral over $k$ through standard contour-integration methods and imposing the normalization condition yields \vspace{-.2cm}
\begin{equation}\label{}
|\varepsilon_b|^2\ug\left(1+\tfrac{\Gamma t_d}{2}\cos[2(k_0+q)x_0]\right)^{-1}\,\,.
\end{equation}
We now replace $\varphi_b(k)$ as given by \eq(\ref{phik2}) in \eq(\ref{epsilon2}), making use again of contour integration, so as to end up with a consistency equation for the rescaled energy parameter
: \begin{equation}
q\ug-\tfrac{\Gamma}{2\upsilon}\sin[2({k}_0+q)x_0].\label{q}
\end{equation}
It is now easy to show that $q\ug0$, i.e., $E\ug\omega_0$. We start by noticing that we can find a value of $k$ that makes the left-hand side of Eq.~\eqref{phik2} vanish, namely $k=k_0+q$. This then entails $\sin[({k}_0+q)x_0]=0$, that is, $({k}_0+q)x_0\ug n\pi$ ($n$ is an integer). This then implies that also the right hand side of Eq.~\eqref{q} vanishes, so that $q=0$ and $\phi\ug2k_0x_0\ug 2n\pi$. 
In conclusion, for $\phi\ug0$ (mod $2\pi$) a bound state $\ket{\Psi_b}$ having energy $E\ug\omega_0$ arises, which significantly overlaps the excited state (the overlap being $\sprod{e\,0}{\Psi_b}=\varepsilon_b$). This explains why full atomic decay to the ground state is inhibited: the projection of the excited state onto the bound state does not couple to the travelling photons, so that, at long times, $\ket{e\,0}\to\ket{\Psi_b}\sprod{\Psi_b}{e\,0}\piu(\rm field\, terms)$ and $\varepsilon(t\!\to\!\infty)=|\varepsilon_b|^2$.
 As is easily checked \cite{nota-confinement}, the amount of atomic excitation corresponding to this overlap remains confined within the interval $0\!\le\!x\!\le\!x_0$, shared between atom and photonic field.
 
This is interpreted as follows. As mentioned, an atom on resonance with a traveling photon behaves as a perfect mirror \cite{shen-fan,switch}. Thus, just like in a standard Fabry-Perot interferometer, one expects a field standing wave to arise within $0\!\le\!x\!\le\!x_0$ when the emitted wavelength matches $x_0$. This agrees with the findings in \rref\cite{sung}, which were however derived in the strong coupling limit. Our $k$-space approach thus allows to prove that a bound state is indeed created and work out explicitly its exact form.

\section{Output field dynamics}  \label{out-field-sec}So far we have focussed on the atomic excitation dynamics. A natural way to experimentally test this is to measure the light emitted through the free -- i.e., non-reflective -- end of the waveguide. It is then important to study the entailed dynamics of such output light. The real-space field annihilation operator at position $x\!>\!0$ can be expressed as
\begin{equation}
\hat C(x)\ug\sqrt{\tfrac{2}{\pi}}\int\!{\rm d}k\,\hat a_k\sin kx,
\end{equation} 
where the pre-factor stems from the normalization constraint $\int_0^\infty {\rm d}x\,\hat C^\dagger(x)\hat C(x)=\int_0^\infty {\rm d}k\,\hat a^\dagger_k\hat a_k$. Once applied to the state in Eq.~\eqref{generic-state} this yields $\hat C(x)\ket{\Psi(t)}=\psi(x,t) \ket g\ket 0$, where
\begin{equation}
\psi(x,t)=\sqrt{\tfrac{2}{\pi}}\int\!{\rm d}k\;\varphi(k,t)\sin kx
\end{equation}
can be interpreted as the real-space field amplitude.
The square modulus of $\psi(x,t)$ can be measured via the local photon density, which is $\propto\!\bra{\Psi(t)}\hat C^\dagger(x)\hat C(x)\ket{\Psi(t)}\ug |\psi(x,t)|^2$.
\begin{figure}
\begin{center}
\includegraphics[width=1\linewidth]{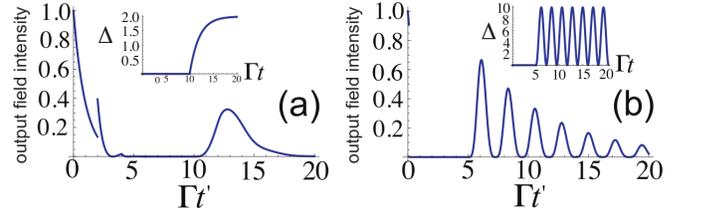}\end{center}\vspace{-.3cm}
\caption{Output field intensity \vs $t'\ug t\meno d/\upsilon$ in arbitrary units for $\phi\ug 2n\pi$ under an applied frequency shift $\Delta(t)$ (this is plotted in the insets in units of $\Gamma$). (a) $\Gamma  t_d\ug 2$ and a steplike $\Delta(t)$. (b) $\Gamma  t_d\ug 0.1$ and a sinusoidal $\Delta(t)$. In either case, a revival of the photon emission occurs when the frequency shift is switched on. Note that, in line with \eq(\ref{out-field1}), at $t\ug t_d$ the intensity exhibits a discontinuity [particularly visible in (a)].
\label{tre}}
\end{figure} 
We assume that a photon detector lies at position $\bar x\ug x_0+d$, where $d\!>\!0$ is the atom--detector distance. Hence (see Appendix \ref{derivation-out-field1}),
\begin{align}
\!\!\!\!\psi(\bar x,t)
&\!=\!\!\!\sqrt{\tfrac{\Gamma}{2\upsilon}}{\rm e}^{ik_0d}[\varepsilon(t')\theta(t')\!-\!{\rm e}^{i\phi}\varepsilon(t'\!-\!t_d)\theta(t'\!-\!t_d)]\label{out-field1}\\
&=\!\!\sqrt{\tfrac{2}{\Gamma \upsilon}}{\rm e}^{ik_0d}\dot\varepsilon(t')\theta(t'),\label{out-field}
\end{align}
where $t'=t-d/\upsilon$, and the last equality follows from Eq.~\eqref{DDE}. Eq.~\eqref{out-field} shows that the time evolution of the atomic excitation (once the time delay $d/\upsilon$ is accounted for) can in fact be obtained by integrating the output field amplitude over time. The latter can be retrieved from the field intensity in the special cases $\phi=0,\pi$, in which the phase of $\varepsilon$ is constant, while in general homodyne techniques will be required. Clearly, for $\phi\ug 2 n\pi$, which ensures the formation of the atom-photon bound state (see previous section), the atom cannot fully decay to the ground state and thus less than one photon exits the waveguide on average. An interesting simple method exists, though, to force the trapped excitation to be released. At a time long enough that the unbound excitation has left the waveguide, an atomic frequency shift $\Delta$ is applied (this is routinely implemented through local fields). As this changes the atomic frequency $\omega_0$ and thereby the corresponding $k_0$, $\phi$ is modified as well. This suppresses the bound state and necessarily compels the trapped excitation to leak out as light. In \fig3, we model the frequency shift switch as a smooth time function $\Delta(t)$ [in a way that in \eq{(\ref{H})} $\omega_0\!\to\!\omega_0+\Delta(t)$, while in Eqs.~\eqref{DDE} and \eqref{out-field} $\dot\varepsilon(t)\!\to\!\dot\varepsilon(t)+i\Delta(t)\varepsilon(t)$] and plot the resulting numerically computed output field intensity against time. Clearly, as soon as $\Delta(t)\!\neq\!0$ a spontaneous  emission revival takes place witnessing that release of the bound state excitation has been triggered.
For $\Gamma t_d\ll1$ [\fig3(b)], the initial emission (when $\Delta$ is still zero) is about negligible because in such regime most of the energy is trapped within the bound state [\cf \eq(\ref{FVT}) and \fig2(b) for $\phi\ug0$]. Interestingly, in such a case the system responds quickly to applied frequency shifts, which can be understood as follows. We start by observing that a nonzero value of $\Delta$ is equivalent to an appropriate phase shift $\phi\to\phi+\delta\phi$. When the phase is shifted from the value $2n\pi$, the bound state is suppressed and the photonic emission revived. From the approximate solution of Eq.~\eqref{eps-app}, one can see that after a transient $\simeq t_d$ the emission rate stabilizes to a fixed value. If later the phase is restored to the bound-state value, Eq.~\eqref{eps-app} again indicates that the system ceases to emit after a further transient time $\simeq t_d$, occurring in a small unwanted excitation loss $\sim\Gamma t_d$. Hence, when $\Gamma t_d\ll1$ these transients have a minor effect so as to allow for a satisfactory degree of control over the atomic emission.
 
As a result, the shape of $\Delta(t)$ is closely reflected in the temporal profile of the light intensity, as shown in Fig.~\ref{tre}(b) in a paradigmatic case. Such effect has the potential to be harnessed to emit single-photon pulses directionally and with controllable temporal profiles, which can be of concern to a variety of fields especially in connection with quantum information technologies. The method outlined here shares some similarities with earlier cavity QED proposals for the deterministic generation of single photons, which typically require the use of more than two internal atomic levels combined with adiabatic transfer techniques \cite{shakespeare}. In contrast, in our setup the existence of a metastable bound state allows for the control of the atomic emission without the need for extra degrees of freedom and through the simple application of a classical field detuning the atom.
 
\section{Resilience to detrimental effects}  \label{resilience}To assess the experimental observability of the central phenomena presented so far, we have refined the model to account for detrimental factors. In addition to the waveguide modes, we allow for an extra atomic coupling to a reservoir of external non-accessible modes at a rate $\Gamma_\text{ext}$. Also, we assume that the guide is terminated at $x=0$ with a non-ideal mirror of reflectivity $R\!<\!1$. Moreover, we introduce (inhomogeneous) phase noise on the atom by adding a small white-noise stochastic term to the excited-state frequency as $\omega(t)=\omega_0\piu\eta(t)$. Here, $\eta(t)$ is a Gaussian-distributed random variable such that $\av {\eta(t)}\ug 0$ and $\av{\eta(t)\eta(t')}=2\delta\omega\,\delta(t\meno t')$ ($\av{\cdot\!\cdot\!\cdot}$ stands for the ensemble average) where $\delta\omega$ represents the associated dephasing rate.
The corresponding excitation probability $P_e(t)\ug\langle|\varepsilon(t)|^2\rangle$ in such non-ideal conditions can be predicted through a semi-analytical procedure (See Appendix \ref{derivation-detrimental}). \\
\begin{figure}
\includegraphics[width=1\linewidth]{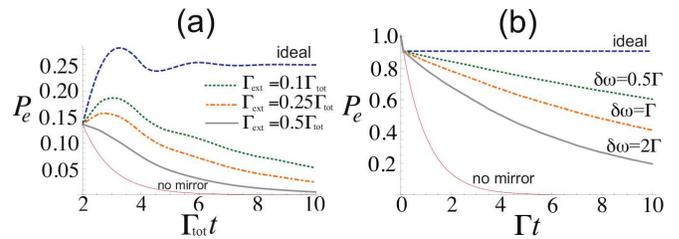}
\caption{{Robustness of $P_e(t)=\langle|\varepsilon(t)|^2\rangle$ \vs time for $\phi\ug 2n\pi$. We have studied the dynamics of the excitation amplitude via Eq.~\eqref{DDE-noise}, using $\Gamma_\text{tot}\equiv\Gamma+\Gamma_\text{ext}$, and $r=R+i\sqrt{R(1-R)}$ . (a) We set $\Gamma_{\rm tot}t_d\ug 2$, $R\ug0.98$, $\delta\omega=0.25\Gamma_\text{tot}$ and vary the ratio between $\Gamma_\text{ext}$ and $\Gamma$, keeping $\Gamma_\text{tot}$ fixed. (b) We fix $R=1,\Gamma_\text{ext}=0$, $\Gamma t_d=0.1$ and vary the dephasing rate $\delta\omega$.} 
\label{vattro}}
\end{figure}
The first photon reabsorption peak of $P_e(t)$ occurring for $\Gamma t_d\!\sim\!1$ [\cf \fig2(a)] is rather robust to dissipation into external modes (expected to be the major detrimental factor affecting such specific feature). As shown in \fig4(a), a `shoulder' is still visible even with $\Gamma_{\rm ext}\ug\Gamma$ (even higher values ensure that SE significantly departs from the mirror-less case). As for the resilience of the bound-state effects, which are stronger for $\Gamma t_d\!\ll\!1$ [\cf\fig2(b)], \fig4(b) shows that in the case of pure dephasing a significantly long-lived excitation trapping still survives for relatively high $\delta \omega/\Gamma$ ratios.
 
\section{Conclusions} \label{conc}
We have investigated the time evolution of spontaneous emission for a two-level system coupled to a semi-infinite 1D photonic waveguide. We have derived an exact delay differential equation for the atomic excitation amplitude. According to this, the atomic excitation undergoes a non-exponential decay which can exhibit oscillations (a signature of partial photon reabsorption) and long time tails. A full decay to the ground state is even inhibited when the emitted wavelength matches the atom-mirror distance, owing to the formation of an atom-photon bound state which we exactly derive. The amount of trapped excitation can be substantial, {and it can be released as a photon by applying a frequency shift to the atom, resulting in a light emission revival}. We have assessed that such phenomena can be observable even in the presence of substantial detrimental effects such as dissipation into unwanted modes and atomic dephasing. This indicates that an experimental demonstration of the key features investigated here may be not far-fetched. We finally point out that an interesting way to regard our system is to consider the mirror as a means to introduce a {\it feedback} mechanism. This ensures that part of the output signal, i.e., the spontaneously emitted light, is re-inserted into the atomic system as input. Significantly, delay differential equations with a similar structure as Equation (6) occur in quantum optics settings with feedback \cite{vittorio}.
\subsection*{Acknowledgements}
{TT and MSK acknowledge support from the NPRP
4-554-1-084 from Qatar National Research Fund. FC acknowledges support from FIRB IDEAS (project RBID08B3FM). We are grateful to V. Giovannetti, B. Garraway, J. Hwang, D. Dorigoni, M. Tame, M. Reimer, A. Nazir, M. Agio, D. Valente and C. Benedetti for useful discussions.

\appendix
\section{Derivation of \eq\eqref{eps-app}}\label{derivation-eps-app}
For $t\leq t_d$, the delay term in \eq(\ref{DDE}) vanishes and thus $\varepsilon(t)={\rm e}^{-{\Gamma}t/2}$. For $t\!>\!t_d$ and $\Gamma t_d\!\ll\!1$, the time delay $t_d$ becomes the shortest time scale in a way that it can be taken as the differential of time. We thus introduce the discrete variable $n$ and, accordingly, define $\varepsilon_n\equiv\varepsilon(nt_d)$. Therefore, $\dot{\varepsilon}\!\simeq(\varepsilon_{n+1}\meno\varepsilon_n)/t_d$, which once replaced in Eq.~\eqref{DDE} gives the recursion indentity
\begin{equation}
\varepsilon_{n+1}=\left(\tfrac{1+{\rm e}^{i\phi}\tfrac{\Gamma t_d}{2}}{1+\tfrac{\Gamma t_d}{2}}\right)\varepsilon_n.
\end{equation}
Using this along with the matching condition at $t\ug t_d$, $\varepsilon_1\ug\varepsilon(t_d)\ug{\rm e}^{-\tfrac{\Gamma t_d}{2}}$, we immediately end up with
\begin{equation}
\varepsilon_{n}=\left(\tfrac{1+{\rm e}^{i\phi}\tfrac{\Gamma t_d}{2}}{1+\tfrac{\Gamma t_d}{2}}\right)^{n-1}\!{\rm e}^{-\tfrac{\Gamma t_d}{2}}.
\end{equation}
By combining the functions for $t\!<\!t_d$ and $t\!>\!t_d$ and reintroducing the continuous time through $n\ug t/t_d$, we find Eq.~\eqref{eps-app} of the main text.
\section{Derivation of Eq.~\eqref{out-field1}}\label{derivation-out-field1}
We start by recalling that the integration in Eq.~(4) of the main text for $\omega_k=\omega_0\piu\upsilon(k-k_0)$ gives
\begin{equation}
\varphi(k,t)=-ig_k\int_0^t{\rm d}s\;{\rm e}^{-i\upsilon(k-k_0)(t-s)}\varepsilon(s),
\end{equation}
where $g_k=\sqrt{\Gamma \upsilon/\pi}\sin{k x_0}$ and the irrelevant phase factor ${\rm e}^{-i\omega_0t}$ has been removed from both functions $\varphi(k,t)$ and $\varepsilon(t)$. Thus, using that the field amplitude in position space is defined as $\psi(x,t)=\sqrt{2/\pi}\int\!{\rm d}k\;\varphi(k,t)\sin kx$, we find
\begin{align}
\psi(x,t)&=\!-i\frac{\sqrt{2\Gamma \upsilon}}{\pi}\!\!\int_0^t\!\!{\rm d}s\!\int\!\!{\rm d}k\sin{(k x_0)}\sin{(k x)}{\rm e}^{-i\upsilon(k-k_0)(t-s)}\varepsilon(s).
\end{align}
The integral over $k$ returns a combination of $\delta$ functions, which makes particularly straightforward the time integration. This yields
\begin{align}
\psi(x,t)\!=\!-i\!\!\sqrt{\tfrac{\Gamma}{2\upsilon}}&\left[{\rm e}^{ik_0(x_0- x)}\varepsilon(t\meno\tfrac{x_0-x}{\upsilon})\theta(x_0\meno x)\theta(\upsilon t\meno x_0\piu x)\right.\nonumber\\
&\left.\!\!+{\rm e}^{ik_0(x- x_0)}\varepsilon(t\meno\tfrac{x- x_0}{\upsilon})\theta(x\meno x_0)\theta(\upsilon t\meno x\piu x_0)\right.\nonumber\\
&\left.\!\!-{\rm e}^{ik_0(x+ x_0)}\varepsilon(t\meno\tfrac{x+ x_0}{\upsilon})\theta(x\piu x_0)\theta(\upsilon t\meno x\meno x_0)\right].
\end{align}
In the special case $\bar x=x_0+d$ ($d>0$) we have
\begin{align}
\psi(\bar x,t)\!=\!-i\sqrt{\tfrac{\Gamma}{2\upsilon}}&{\rm e}^{ik_0d}\left[\varepsilon(t')\theta(t')\!-\!{\rm e}^{i\phi}\varepsilon(t'\!-\!t_d)\theta(t'\!-\!t_d)\right],\label{crampo}
\end{align}
where $t'=t\!-\!d/\upsilon$ as in the main text. Eq.~\eqref{crampo} is equivalent to Eq.~\eqref{out-field1} of the main text, up to an irrelevant phase factor $-i$.
\section{Including detrimental effects}\label{derivation-detrimental}
Here, we briefly explain how we have extended our model so as to include losses and perform the robustness study in Fig.~4 of the main text. For our purposes, it suffices to adopt a heuristic reasoning (a more rigorous analysis yields the same results). The inclusion of an external reservoir of non-accessible modes is a routine procedure in the literature. It simply amounts to adding a term $-\tfrac{\Gamma_\text{ext}}{2}\varepsilon(t)$ on the right-hand side of \eq\eqref{DDE}, where $\Gamma_\text{ext}$ is the decay rate associated to such unwanted modes. The presence of an imperfect mirror with $R\!<\!1$ will instead modify the delay term in Eq.~\eqref{DDE} since the atom will re-interact only with the portion of light which is reflected. This suggests the substitution ${\rm e}^{i\phi}\to r{\rm e}^{i\phi}$, where $-r$ is the complex probability amplitude for backwards reflection off the mirror ($|r|^2=R$). {Solving the 1D scattering problem yields $r=R+i\sqrt{R(1-R)}$}. Finally, as mentioned in the main text, we include extra dephasing of the atom by adding a white-noise term to the excited-state frequency as $\omega(t)=\omega_0\piu\eta(t)$, where $\eta(t)$ is a Gaussian-distributed random variable: $\av {\eta(t)}\ug 0$ and $\av{\eta(t)\eta(t')}=2\delta\omega\,\delta(t- t')$, where $\delta\omega$ quantifies the strength of the dephasing [$\av{\cdot\!\cdot\!\cdot}$ stands for the ensemble average]. In conclusion, Eq.~\eqref{DDE} of the main text is modified as follows:
\begin{align}
\!\!\dot\varepsilon(t)\!+\!i\eta(t)\varepsilon(t)\ug\!-\frac{\Gamma\piu\Gamma_\text{ext}}{2}\varepsilon(t)\piu r\frac{\Gamma}{2}\, {\rm e}^{i\phi}\!\varepsilon(t\!-\! t_d)\theta(t\!-\! t_d).\label{DDE-noise}
\end{align}
For completeness, we also mention that a similar analysis can be carried out on the output field amplitude, which modifies Eq.~\eqref{out-field1} of the main text as
\begin{equation}
\!\!\psi(\bar x,t)
\!=\!\!\sqrt{\tfrac{\Gamma}{2\upsilon}}{\rm e}^{ik_0d}[\varepsilon(t')\theta(t')\!-\!r{\rm e}^{i\phi}\varepsilon(t'\!-\!t_d)\theta(t'\!-\!t_d)].\!\!\label{out-field-realistic}
\end{equation}
To obtain each line in Fig.~4 of the main text, we have integrated Eq.~\eqref{DDE-noise} numerically for 100 realizations in the presence of simulated white noise, and then averaged over these the resulting probabilities $P_e(t)=|\varepsilon(t)|^2$.
 
Finally, let us stress again that Eqs.\eqref{DDE-noise} and \eqref{out-field-realistic} can be obtained rigorously by modifying the microscopic model given by Eq.~\eqref{H} of the main text. In particular, one has to include both sine and cosine waves for each wavevector $k$, while the presence of an imperfect mirror at $x=0$ can be modeled by adding an extra term in the Hamiltonian of the form $V\hat{C}^\dagger(0)\hat{C}(0)$. Here, $V=\upsilon\sqrt{R/(1-R)}$ and $\hat{C}(x)$ is the field annihilation operator in real space as introduced in the main text [now however owing to the cosine standing modes $\hat{C}(0)\!\neq\!0$].


\begin {thebibliography}{99}
\bibitem{purcell} E. M. Purcell, Phys. Rev. {\bf 69}, 681 (1946).
\bibitem{meschede} D. Meschede, Phys. Rep. {\bf 211}, 201 (1992).
\bibitem{cqed} H. Walther, B. T. H. Varcoe, B.-G. Englert, and T. Becker, Rep. Prog. Phys. {\bf 69}, 1325 (2006); R Miller \etal, J. Phys. B: At. Mol. Opt. Phys. {\bf 38}, S551 (2005); J. M. Raimond, M. Brune, and S. Haroche, Rev. Mod. Phys.
{\bf 73}, 565 (2001).
\bibitem{bandgap} S. John and J. Wang, Phys. Rev. Lett. {\bf 64}, 2418 (1990); S. Bay,
P. Lambropoulos, and K. Moelmer, Phys Rev Lett {\bf 79}, 2654 (1997).
\bibitem{pc} A. Faraon {\it et al.}, \apl {\bf 90}, 073102 (2007).
\bibitem{fibers} B. Dayan {\it et al.}, Science {\bf 319}, 1062 ( 2008); E. Vetsch {\it et al.}, \prl {\bf 104}, 203603 (2010); M. Bajcsy {\it et al.}, \ibid {\bf 102},  203902 (2009).
\bibitem{wallraf} A. Wallraff et al., Nature (London) {\bf 431}, 162 (2004); O. Astafiev {\it et al.} Science {\bf 327}, 840 ( 2010).
\bibitem{nws} M. H. M. van Weert {\it et al.}, Nano Lett. {\bf 9}, 1989 (2009); T. M. Babinec {\it et al.}, Nat.
Nanotechnol. {\bf  5}, 195 (2010).
\bibitem{gerard} J. Claudon {\it et al.},  Nature Photonics {\bf 4}, 174 (2010); J. Bleuse {\it el.}, \prl {\bf  106}, 103601 (2011).
\bibitem{delft} {M. E. Reimer {\it et al.}, Nat. Commun. {\bf 3}, 737 (2012); G. Bulgarini {\it et al.}, Appl. Phys. Lett. {\bf 100}, 121106 (2012).}
\bibitem{plasmons} A. Akimov A. {\it et al.}, Nature {\bf 450}, 402 (2007); A. Huck, S. Kumar, A. Shakoor and U. L. Andersen, Phys. Rev. Lett., {\bf 106}, 096801 (2011).
\bibitem{sorensen} D. Witthaut, and A. S. S\o rensen, New J. Phys. {\bf 12}, 043052 (2010).
\bibitem{freespace} G. Zumofen, N. M. Mojarad, V. Sandoghdar, and M. Agio, Phys. Rev. Lett., {\bf 101}, 180404 (2008);  N. Lindlein, R. Maiwald, H. Konermann, M. Sondermann, U. Peschel,
G. Leuchs, Las. Phys. {\bf 17}, 927 (2007).
\bibitem{agnellone} P. Horak, P. Domokos and H. Ritsch,
Europhys Lett {\bf61}, 459, (2003).
\bibitem{shen-fan} J.-T. Shen and S. Fan,  Opt. Lett. {\bf 30}, 2001 (2005); Phys. Rev. Lett. \textbf{95},
213001 (2005).
\bibitem{lukin} D. E. Chang, A. S. S\o rensen A. S., E. A. Demler, and M. D. Lukin, Nat. Phys. {\bf 3}, 807 (2007).
\bibitem{switch} L. Zhou {\it et al.}, Phys. Rev. Lett. {\bf 101}, 100501 (2008).
\bibitem{friedler} I. Friedler \etal, Opt. Express {\bf 17}, 2095 (2009).
\bibitem{jap} A. V. Maslov, M. I. Bukanov, and C. Z. Ning, J. Appl. Phys. 
{\bf 99}, 024314 (2006).
\bibitem{sung} H. Dong {\it et al.}, Phys. Rev. A {\bf 79}, 063847 (2009).
\bibitem{solano} B. Peropadre {\it et al.}, Phys. Rev A {\bf 84}, 063834 (2011).
\bibitem{chen} Y. Chen, M. Wubs, J. M\o rk and A. F. Koenderink, New J. Phys. {\bf 13}, 103010  (2011).
\bibitem{gate} F. Ciccarello {\it et al.},  Phys. Rev A {\bf 85}, 050305(R) (2012).
\bibitem{valente} D. Valente, Y. Li, J. P. Poizat, J. M. Gerard, L. C. Kwek, M. F. Santos, and A. Auffeves, New J. Phys. {\bf 14}, 083029 (2012); D. Valente, Y. Li, J. P. Poizat, J. M. Gerard, L. C. Kwek, M. F. Santos, and A. Auffeves, Phys. Rev. A {\bf 86}, 022333 (2012).
\bibitem{supercavity} L. Zhou {\it et al.}, Phys. Rev. A 
{\bf 78}, 063827 (2008).
\bibitem{dde} Ho Trung Dung and Kikuo Ujihara, Phys Rev A {\bf59}, 2524 (1999).
\bibitem{dde2} Fam Le Kien, Nguyen Hong Quang and K. Hakuta , Opt Commun {\bf 178}, 151 (2000).
\bibitem{nota-1order} As expected, \eq(\ref{FVT}) agrees with \eq(\ref{eps-app}) at first order in $\Gamma t_d$. We stress, however, that the result in \eq(\ref{FVT}) is exact and thus valid for any $\Gamma t_d$.
\bibitem{nonmarkovianeffect} Qing-Jun Tong, Jun-Hong An, Hong-Gang Luo, C. H. Oh, J. Phys. B \textbf{43}, 155501 (2010).
\bibitem{nota-confinement} This is easily seen by calculating the real space amplitude $\psi_b(x)\propto\int{\rm d}k\; \varphi_b(k)\sin(kx)$ (see {\it Output field dynamics}). Standard contour integration methods, together with the condition $k_0x_0=n\pi$, yield $\psi_b(x)\propto\sin(k_0x)\theta(x_0\meno x)$, hence showing that the field amplitude is exactly zero for $x>x_0$.
\bibitem{shakespeare} C. K. Law and  H. J. Kimble, J. Mod.
Opt. {\bf44}, 2067 (1997); A. Kuhn, M. Hennrich, T. Bondo and G. Rempe, Appl. Phys. B {\bf69}, 373 (1999).
\bibitem{vittorio} V. Giovannetti, P. Tombesi and D. Vitali, Phys. Rev. A {\bf60}, 1549 (1999).
 
\end{thebibliography}
\end{document}